# A Multi QoS Genetic-Based Adaptive Routing In Wireless Mesh Networks With Pareto Solutions

**Ibraheem Kasim Ibraheem**
Department of Electrical Engineering, Baghdad University, Baghdad, Iraq
Email: ibraheemki@coeng.uobaghdad.edu.iq

**Alyaa Abdul-Hussain Al-Hussainy**
Department of Electrical Engineering, Baghdad University, Baghdad, Iraq
Email: alyaaalhussainy79@gmail.com

*Abstract*—Wireless Mesh Networks(WMN) is an active research topic for wireless networks designers and researchers. Routing has been studied in the last two decades in the field of optimization due to various applications in WMN. In this paper, Adaptive Genetic Algorithm (AGA) for identifying the shortest path in WMN satisfying multi- *QoS* measure is introduced. The proposed algorithm is adaptive in the sense that it uses various selection methods during the reproduction process and the one with the best multi- *QoS* measure is adopted in that generation .The multi-objective *QoS* measure defined as the combination of the minimum number of hops, minimum delay, and maximum bandwidth. The multi-objective optimization has been formulated and solved using weighted sum approach and Pareto optimal solution techniques. The simulation experiments have been carried out in MATLAB environment with a wireless network modeled as weighted graph of fifty nodes and node coverage equals to 200 meter, and the outcomes demonstrated that the proposed AGA performs well and finds the shortest route of the WMN proficiently, rapidly, and adapts to the dynamic nature of the wireless network and satisfying all of the constraints and objective measures imposed on the networks.

*Index Terms*— **Quality-of-Service (*QoS*), wireless routing, end-to-end delay, network bandwidth, wireless mesh networks, number of hops.**

## I. Introduction

The routing in multi-hop mobile networks has been considered one of the outstanding design matters. Which significantly affects their attainable achievement. Subsequently, proficient routing methods ought to be intended for guaranteeing that the information packets proliferate in an "ideal" way regarding a few measures, for example, packet loss ratio, defer-jitter, delay, and bandwidth. All the preferred goals are improved all together in the conventional multi-hop mobile networks. However, in certain commonsense applications, to find the different solutions, each of which is ideal regarding an individual *QoS* measure might be superior to finding a solitary worthy arrangement, which incurs a balance among a few different variables [1]. A few Shortest Path (SP) search algorithms like Bellman-Ford and Dijkstra perform successfully for settled framework wireless or wired networks. In any case, they experience the ill effects of high computational complexities in networks with quickly changing topology as well as system status. Classically, SP routing problem has been formulated to combinatorial optimization that seeks to find the single best solution in one run. Routing in mobile networks involves simultaneous optimization of multiple *QoS* parameters such as the delay, the bandwidth, hops number, losses, etc. These objectives compete and conflict with each other. Such competition among conflicting objectives gives rise to a set of optimal solutions instead of a single solution [2].

This paper is organized as follows. Related work is presented in section II. Section III introduces a concise introduction to the GA. Multi-Objective optimization and Pareto Solutions are reviewed in section IV. In section V, a detailed description of the proposed AGA for routing in WMN is presented and discussed. The effectiveness of our proposed algorithm and the discussion of the simulation results are given in section VI. Finally, the conclusions are mentioned in section VII.

## II. Related Works

Many works have been focused on routing in WMN. In [3], authors Proposed a multi-objective traffic engineering procedure utilizing various distribution trees to several multicasting flows. The purpose is to combine into a single united measure the bandwidth, hop count, supreme link utilization, and total delay. The work in [4] studied the performance of several algorithms for multi-objective Pareto optimization. These algorithms were tested on a set of standard benchmark problems. Where in [5] researchers proposed a novel technique for resembling the Pareto front of a Multi-Objective Optimization (MOP) problem, where explicit forms of the objective functions are not available. The method iteratively approximates each objective function using a meta-modeling scheme and employs a weighted sum method to convert the MOP into a set of single objective optimization problems. A work utilizing GA and multi-objective optimization for *QoS* Routing in Wireless Ad-hoc Networks was proposed in [6]. To limit the searching field of the GA, a technique for reducing the searching has been implemented, which reduced the search space to find a new route. A new alternate for the routing problem in WMN has been presented, which considers the *QoS* measure [7].

The actual case study includes several design objectives which conflict with each other. The new approach is tried to improve the routing solutions and proposes the use of Multi-Objective Evolutionary Algorithms (MOEA), specifically the Nondominated Sorting GA (NSGA). A mathematical model is introduced for this problem, which includes *QoS* parameters such as bandwidth, packet loss rates, and delay and power consumption. Jitter mechanisms can dramatically improve reactive routing protocols, in [8], jitter mechanisms are proposed which enforce wireless nodes to postpone their transmission for a random amount of time so as to reduce probability of simultaneous transmission. The work in [9] paper proposed a centralized MultiPAth *QoS*-driven Routing (MPAR) protocol for industrial WMN which included the end-to-end reliability requirements of the available paths. On the other hand, stability of wireless mesh networks is an important issue; instability in these networks is caused mainly by link quality fluctuations and frequent route flapping.

Authors in [10] addressed the stability problem of wireless mesh networks. A routing protocol which applied Software Defined Networks (SDN) to multi-hop wireless network has been studied in [11]. The proposed protocol is implemented using OPNET simulation. For Hybrid WMNs, [12] proposed a load-aware cooperative hybrid routing protocol (LA-CHRP). LA-CHRP is not only adapted to cover the peculiarities of routers and clients, but also considers load in routing. Finally, a GA-based Multi-Path QoS Routing (MPQR) scheme is proposed for Polar-orbit Low earth orbit satellite networks [13].

The work presented in this paper is an extension for our previous works in [14–17]. In this paper, a new approach called Adaptive Genetic Algorithm (AGA) has been proposed by implementing the reproduction process with variable selection methods. The algorithm called adaptive because it chooses the selection method adaptively according to the best value of the fitness function that each of the six selections methods produces. Then the proposed Adaptive GA has been applied to obtain the shortest route in wireless networks where the nodes of the wireless networks are mobile with time.

## III. A GENETIC ALGORITHMS (GA)

GAs are an evolutionary optimization approach, they are especially appropriate for applications which are vast, nonlinear and potentially discrete in nature. In GA, a population of strings called chromosomes (or individuals) which represent the candidate solutions to an optimization problem is evolved to the better population. It is more common to state the objective of GA as the minimization of some cost function rather than the maximization of some utility or fitness function [18],

$$F(t) = \frac{1}{1+f(t)} \quad (1)$$

Where *f(t)* is the cost function to be maximized. While the fitness is set to $F(t) = f(t)$ when the problem needs *f(t)* to be minimized. GA consists of main thee steps, these are: selection process, crossover, and mutation. The Selection process refers to the mechanism of choosing a set of chromosomes from the population that will contribute to the creation of the offsprings for the next generation [19]. Many methods have been proposed for mate selection in the literature, some of the important methods are described in our previous works [17,20], these include Roulette Wheel Selection (RWS), Tournament selection (TS), Rank selection (RS), Steady-state selection (SSS), Sigma scaling selection (SigSS) and Boltzmann selection (BS). On the other hand, crossover operation, or mating, is the creation of one or more offspring from the parents selected in the pairing process. The final step of the GA is the mutation operator, it is another way of the GA to investigate the cost-surface and it can introduce individualities that never exists in the principal population and preserve the GA from converging too fast before searching the complete cost-surface [21,22].

## IV. B PARETO APPROACH FOR MULTI-OBJECTIVE OPTIMIZATION

For sound-organized mathematical problems, the weighted sum (WS) strategy achieves well and has decent scientific properties, for example, convergence. Nonetheless, the WS technique cannot create any point in the non-convex region of the Pareto front as illustrated in Fig.1. Point C, D, and E are not on the Pareto frontier because it is dominated by both point A and point B. Points A and B are not strictly dominated by any other, and hence do lie on the frontier. Additionally, WS technique may copy solutions with various weighting factors [5].

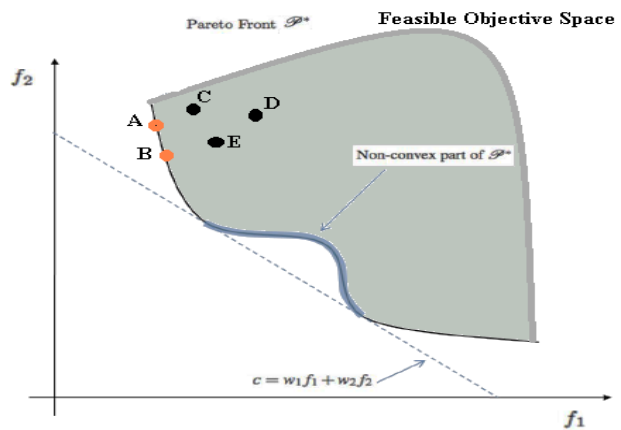

Fig. WS is unable to generate the non-convex part of the Pareto front.

The second general approach for multi-objective optimization is to find whole Pareto optimal solutions or a typical subset. A Pareto optimal solution is a set of solutions that are non-dominated with each other. While traveling from one Pareto solution onto the next, there must be a sacrifice in one of the objective(s) to accomplish a specific improvement in the other(s) [23–25]. Genetic algorithms (GAs) work with a population of points, a number of Pareto-optimal solutions may be captured using GAs. A new algorithm called Nondominated Sorting Genetic Algorithm (NSGA) is presented in this work. This algorithm eliminates the bias in Vector Evaluated Genetic Algorithm (VEGA) [24,25] and thereby distributes the population over the entire Pareto-optimal regions [26,27]. In this work three objectives: an end-to-end delay, bandwidth, and a number of hops will be utilized as multi-objective measures using NSGA algorithm to test their convergence through the generations of the GA.

Definition 1 [28]: A solution $x^{(1)}$ is said to dominate the other solution $x^{(2)}$ if both conditions 1 and 2 are true:

1. The solution $x^{(1)}$ is no worse than $x^{(2)}$ in all objectives: that is,

$$f_j(x^{(1)}) \triangleright f_i(x^{(2)}) \quad \text{for all } j = 1, 2, 3, \ldots, M$$

2. The solution $x^{(1)}$ is strictly less than $x^{(2)}$ in at least one objective, or

$$f_{\bar{j}}(x^{(1)}) \triangleleft f_{\bar{j}}(x^{(2)}) \quad \text{for at least one } \bar{j} \in \{1, 2, \ldots, M\}$$

Operator $\triangleleft$ between two solutions $i$ and $j$ as $\triangleleft$ to denote that solution $i$ is better than solution $j$ on a particular objective. Similarly, $i \triangleright j$ for a particular objective implies that solution $i$ is worse than solution $j$ on this objective. If either of the above conditions is violated, the solution $x^{(1)}$ does not dominate the solution $x^{(2)}$. If $x^{(1)}$ dominates the solution $x^{(2)}$ (or mathematically $x^{(1)} \preccurlyeq x^{(2)}$), it is also customary to write any of the following:

- $x^{(2)}$ is dominated by $x^{(2)}$,
- $x^{(1)}$ is non-dominated by $x^{(2)}$, or $x^{(1)}$ is non-inferior to $x^{(2)}$

The set of all feasible solutions that are non-dominated by any other solution is called the Pareto-optimal or non-dominated set. If the non-dominated set is within the entire feasible search space, it is called globally Pareto-optimal set. In other words, for a given MOOP, the Pareto-optimal set $P^*$, is defined as:

$$P^* = \{x \in \Omega | \neg \exists\, x' \in \Omega \Rightarrow f(x') \leq f(x)\}$$

The values of objective functions related to each solution of a Pareto-optimal set in objective space are called Pareto-front. In another word for a given MOP, f(x), and Pareto-optimal set, PF∗, the Pareto front( PF∗) is given by:

$$PF^* = \{u = f(x) | x \in P^*\}$$

V.  THE PROPOSED AGA FOR ROUTING IN WMN

The proposed AGA consists of the following steps:

A. *Priority-Based Encoding*

When utilizing GA based routing in wireless networks, each candidate solution called a chromosome. The chromosome consists of sequences of positive integers that represent the IDentification (ID) of nodes through which a routing path passes. Each locus of the chromosome represents an order of a node in a routing path. A certain gene in a specific chromosome is described by the variables: *allele*, the value for that gene, and loci, the gene's position. The gene's position is utilized to characterize a node, while the priority for that node for developing a path from competitors is represented by the value of the gene. This strategy is signified as Priority Based Encoding as illustrated below (see Fig. 2).

| Position: node | 1 | 2 | 3 | 4 | 5 | 6 | 7 | 8 | 9 | 10 |
|---|---|---|---|---|---|---|---|---|---|---|
| Value: priority | 7 | 3 | 4 | 6 | 2 | 5 | 8 | 1 | 1 | 9 |

Fig 2.  Representation of Priority Based Encoding.

B. *Representation*

In this case-study, the wireless routing scheme is represented as the GA chromosome. A route list interprets the routing chromosome, such as P = (P1, P2 …, Pk), which characterizes the complete network. Every route is a particular path between the source and destination nodes i and j. A route is coded by concatenating the nodes from the start node to the target node depending on the network topology. For example, a route starting from node number one to node number fifty can be denoted as a node vector along the route: {1, 2, 4, 12, 34, 50}, see Fig. 3. If a route cannot be achieved on the network, it cannot be coded into the chromosome.

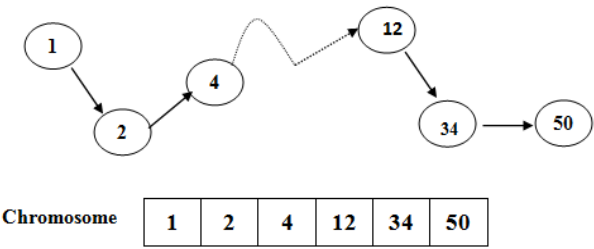

Fig 3.  Example of routing path and its encoding scheme.

C. *Initial Populations*

This underlying procedure is utilized to create the routing table in the present generation. Every chromosome incorporates an arbitrary routing table for the particular topology of the mobile network under study.

D. *Multi-Objective Optimization*

Different *QoS* criteria characterize routing in mobile wireless meshed networks, some of these criteria are a number of hops, an end-to-end delay, and bandwidth which are adopted in this work. Clearly, it can be fit into multi-objective optimization. In this work, we propose the design of a Multi-Objective GA( MOGA) through the Weighted Sum (WS) approach. This method is used in MOGA to attain the above three objectives through a single-objective measure by utilizing the convex combination of the design measures. Minimizing the fitness function means minimization of the weighted sum function $F$ given by the following formula:

$$F = \alpha_1 F_1 + \alpha_2 F_2 + \alpha_3 F_3 \qquad (2)$$

Where $F_1$, $\acute{F}_2$, and $F_3$ are the delay, the bandwidth, and the number of hops, given by,

$$F_1 = \min\ \text{Delay}(P(s,d)) \tag{3}$$

$$\bar{F}_2 = \max\ \text{Bandwidth}(P(s,d)) \tag{4}$$

$$F_2(t) = \frac{1}{1+\bar{F}_2} \tag{5}$$

$$F_3 = \min\ \text{Hop}(P(s,d)) \tag{6}$$

Subject to

$$\text{Delay}(P(s,d)) <= Dmax$$

$$\text{Bandwidth}(P(s,d)) >= Bmin$$

$$\text{Hops}(P(s,d)) <= Hopsmax$$

Where $P(s,d) = P_1, \ldots\ldots P_n$, is the collection of all the potential routes from source node $s$ to destination node $d$, where a certain source-node $s \in V$ and destination-node $d \in V$, $V$ is the network nodes set (vertices), $\text{Delay}(P(s,d)) = \sum \text{Delay}(V)$ is the total delay along the entire path $P_1, \ldots\ldots P_n$, $\text{Bandwidth}(P(s,d)) = \max \text{Bandwidth}(V)$ is the maximum bandwidth along the entire path $P_1, \ldots\ldots P_n$, $\text{Hops}(P(s,d)) = \sum \text{Hops}(V)$ is the total number of hops along the entire path $P_1, \ldots\ldots P_n$, $Dmax$ represents the maximum limit on end-to-end delay along each path, $Bmin$ represents the lower bound on the acceptable bandwidth along each path from a source s to the destination $d$, $Hopsmax$ represents the upper bound on the acceptable number of hops along every route from the start $s$ to the terminating node $d$. the weights $\alpha_1$, $\alpha_2$, $\alpha_3$ are understood as the relative importance of one objective function relative to the others. The values of $\alpha_1$, $\alpha_2$, $\alpha_3$ are selected to increase the selection weight on any of the three objectives, such that,

$$\sum_{i=1}^{n} \alpha_i = 1 \tag{7}$$

*E. Proposed Selection Process*

Selection is an operator to select two (i.e., routing tables) for generating new chromosomes. To be selected as a parent chromosome, the chromosomes are competing each other during selection process based on the fitness value. Each chromosome has its fitness value calculated according to (2), more explanations on different selection operators used in this work can be found in our previous works [17,20]. A chromosome with a high fitness value (*i.e.* minimum cost) subject to the delay, maximum bandwidth, and minimum number of hops has more chance to be selected as one of the parents using one of the six selection methods that had the minimum fitness value.

*F. Path Crossover*

To carry out the crossover operator, the chromosomes must have the same start and target nodes. The points in the chromosomes are constrained to the nodes that are common in both chromosomes. The most common and the simplest form of crossover is single point crossover. A crossover site is randomly selected on both chromosomes and exchanging the sub-routes when carrying out this operator to both chromosomes as depicted in Fig. 4.

Selecting node number 11 as a cross-node results in the child pair of chromosomes given as $P_1$' and $P_2$'. The Procedure of the path crossover can be summarized in the following:

1. Create list of nodes *NC* which exist in both $P_1$ and $P_2$ (excluding source (s) and destination (d) nodes) as possible a cross-point.
2. From the list *NC*, pick a certain node *i* as a cross-point.
3. Exchange all the nodes for the parent chromosomes beyond the cross-point *i* to achieve the crossover process.

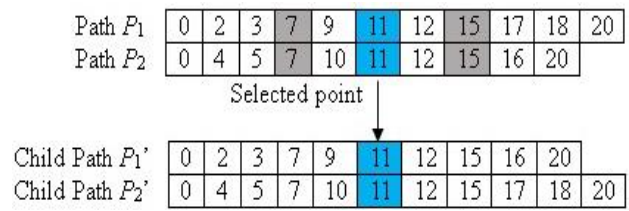

Fig. 4. An example of Single point Crossover.

*G. Path Mutation*

The operator of the route mutation means generating another chromosome from the chromosome. To apply the mutation operator, firstly a single node is arbitrarily chosen from each chromosome, it is denoted the mutation's site. At that point, another node is arbitrarily chosen from the set of nodes that are directly connected to the mutation site. In accordance with the shortest path problem, a second path is determined through linking starting node to chosen mutation node and from the chosen node to target node. As illustrated in Fig. 5, where the offspring P' signifies the route obtained by the mutation operator. The path mutation operator procedure is described as follows:

1. From all nodes in parent *P*, randomly pick a node *i* as a mutation node.
2. From the neighbors (*B*) of the mutation node *i*, select a node *j* ∈ *B*.
3. Create two paths *r*1 and *r*2 with shortest distance, the first (r1) from the mutation node *j* to the source node (*s*) and the second(*r*2) from the destination node (*d*) to node *j*.
4. If nodes replication occurs between shortest paths (*r*1 and *r*2), cancel the paths and stop mutation, else join the path to create a mutation node.

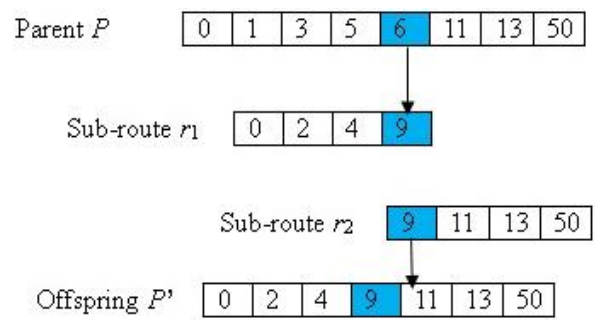

Fig. 5. An example of Mutation.

## VI. MAIN RESULTS

### A. Initialization

A network model is obtained by $G$ ($V$, $E$), where $G$ is a graph and V is the collection of the nodes of the network, and E is the collection of the connected linked edges. The simulation network parameters are chosen as follows: Number of nodes = 50, as shown in Fig. 6. Topology-area: Nodes are distributed randomly on 1000*1000 $m^2$, the area of the node coverage equals to 200 $m$, node 1 is the source node (s), while the destination node (d) is node 50. Population size equals to 50, selection method used in this project are {RS, SSS, BS, RWS, SigSS, TS}, number of generations equals to 100, the crossover probability equals is $Pc$ = 0.75, mutation probability is $Pm$ = 0.01. We have fifty routes from s to d, and each route has a fitness number depends on its delay, bandwidth, and number of hops from s to d. We have to find one optimum path among the fifty paths (50 paths) available with minimum delay, maximum bandwidth, and minimum no. of hops. The values of $\alpha_1$, $\alpha_2$, $\alpha_3$ are 0.5, 0.15, 0.35 respectively.

| Selection method Index | Selection method | Maximum | Minimum |
|---|---|---|---|
| 1 | RWS | 15.28 | 11.92 |
| 2 | TS | 15.46 | 11.62 |
| 3 | SSS | 15.36 | 11.92 |
| 4 | BS | 15.42 | 11.92 |
| 5 | Sig SS | 15.42 | 11.92 |
| 6 | RS | 16.15 | 11.92 |

TABLE 1. SELECTION METHOD OF MULTI-OBJECTIVES QOS

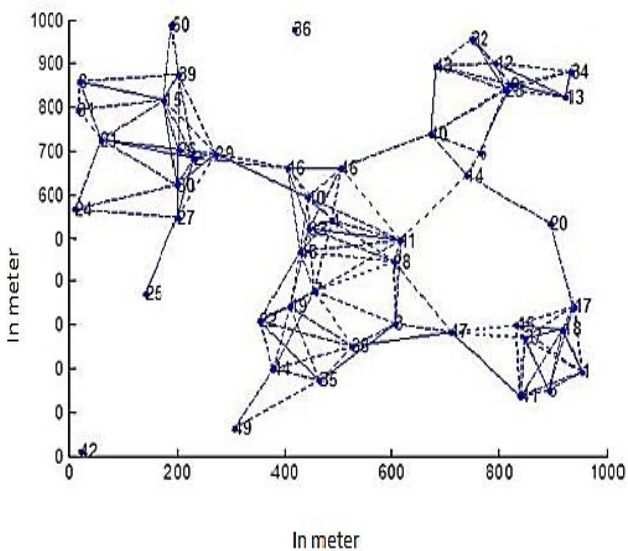

Fig. 6. The network model of 50 nodes

### B. Proposed Selection Methods

The values of the fitness functions with each selection technique are represented by their minimum and maximum values, as illustrated in Table 1. They are applied at the same time. Our experiment concerns about the maximum value of the path bandwidths, minimum values of end-to-end path delays, and minimum path hops.

### C. Optimization Results for the Number of Bandwidth, hops, and End-to-End Delay

The outcomes of implementing our proposed AGA are illustrated as next, the shortest path from the starting node (s) to the destination node (d) was {1 24 26 46 50} as depicted in Fig. 7. The end-to-end delay is {8 msec}, bandwidth was 1.9932 Mbps and a number of hops was 4. The best fitness value was (10.2968) resulted from selection method 3, namely, the SSS selection method. Therefore, the SSS selection method gives the best fitness. Fig. 8 showed the variation of the fitness values during iterations. Additional results for the multi-objective $QoS$ experiment are given in in Figs. 9 and 10, where Fig. 9 shows the values of the $QoS$ parameters of some paths, $P_1$, $P_2$, $P_3$, $P_4$. It can be concluded that P4 is the best optimum path among them. Fig. 10 showed the fitness values for the multi-objective $QoS$ for various paths, $P_1$, $P_2$, $P_3$, $P_4$ knowing that path four P4 is the optimum path. To solve multi-objective optimization using NSGA with Pareto solution approach, two main goals must be attained. The first goal is to converge to a set of solution as close as possible to true Pareto-optimal set, and the second goal is the diversity in the obtained Pareto-optimal set. With a more diverse set of solutions that covers all parts of the Pareto-front in objective space, the decision making process at the next level using the higher level information is easier. The diversity in the two-dimensional space is often Symmetric, however in three -dimensional space (three objectives problem) and non-linear problem the diversity is more difficult to obtain. Fig.11 (a) illustrated the Pareto optimal solutions of the three objectives optimization, where $f(x_1)$, $f(x_2)$, $f(x_3)$ represent end-to-end delay, bandwidth, and number of hops respectively. The population size is 50 and the number of generations are 100, 200, 4000, and 10000. The diversity and the convergence of the Pareto solutions are very obvious. The weighed-sum approach represents the Pareto solution to multi-objective optimization; this was clear from Fig. 12.

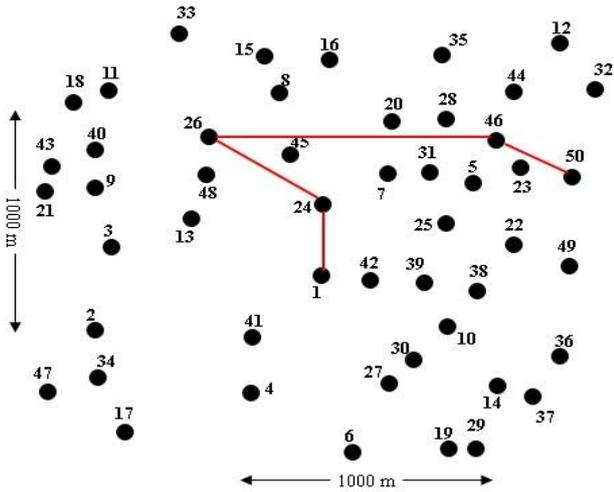

Fig. 7 Network Topology is showing the optimal path (red line) satisfying multi-objectives QoS measures, delay, bandwidth, and a number of hops.

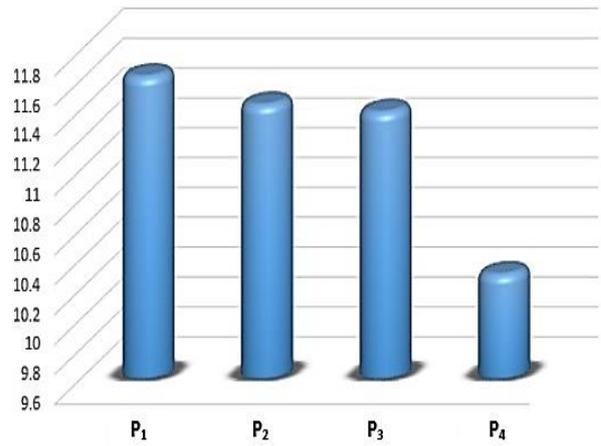

Fig. 10 Fitness value of Multi-objective parameters for four paths.

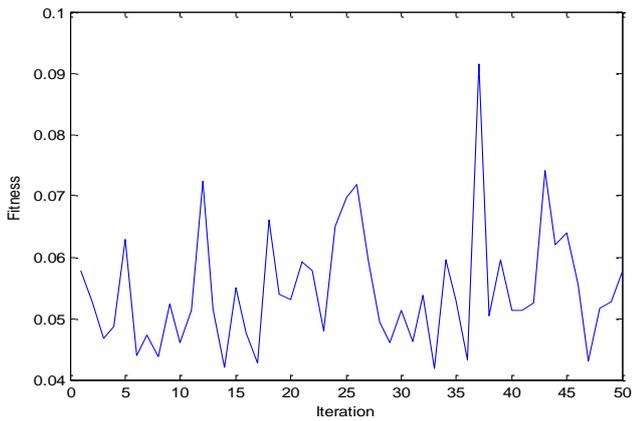

Fig. 8 Normalized Fitness variation through iterations for multi-objective QoS.

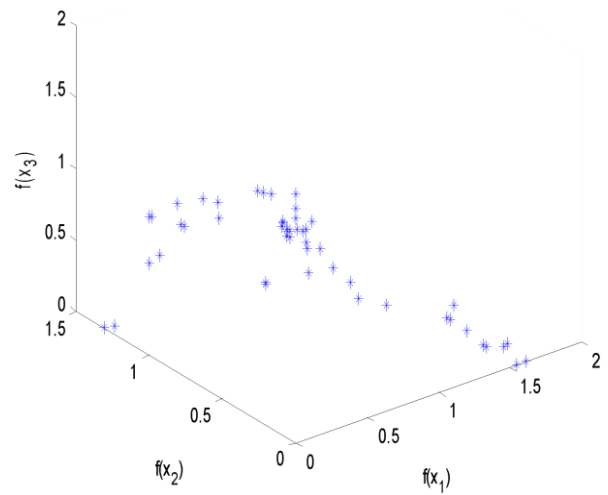

(a) 100 generations

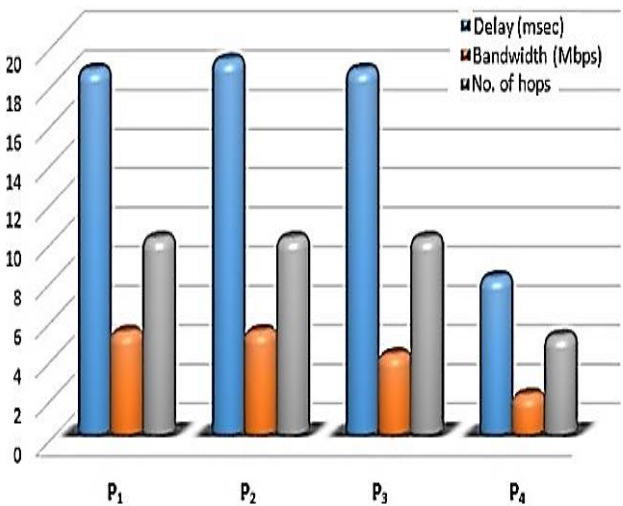

Fig. 9 Multi-objective QoS parameters for the four paths.

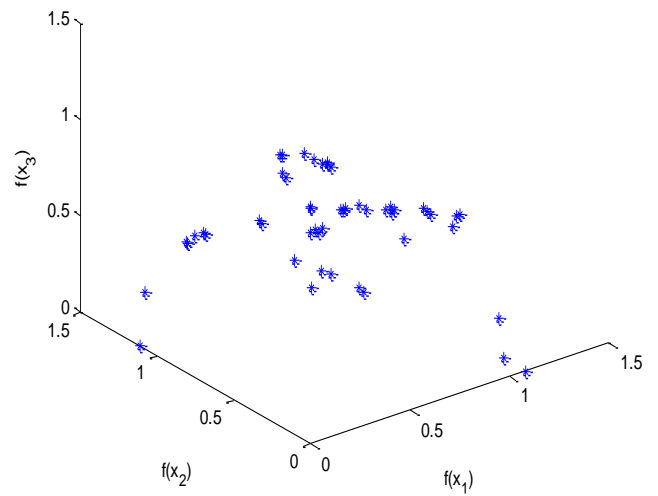

(b) 200 generations

Fig. 11 Continued…

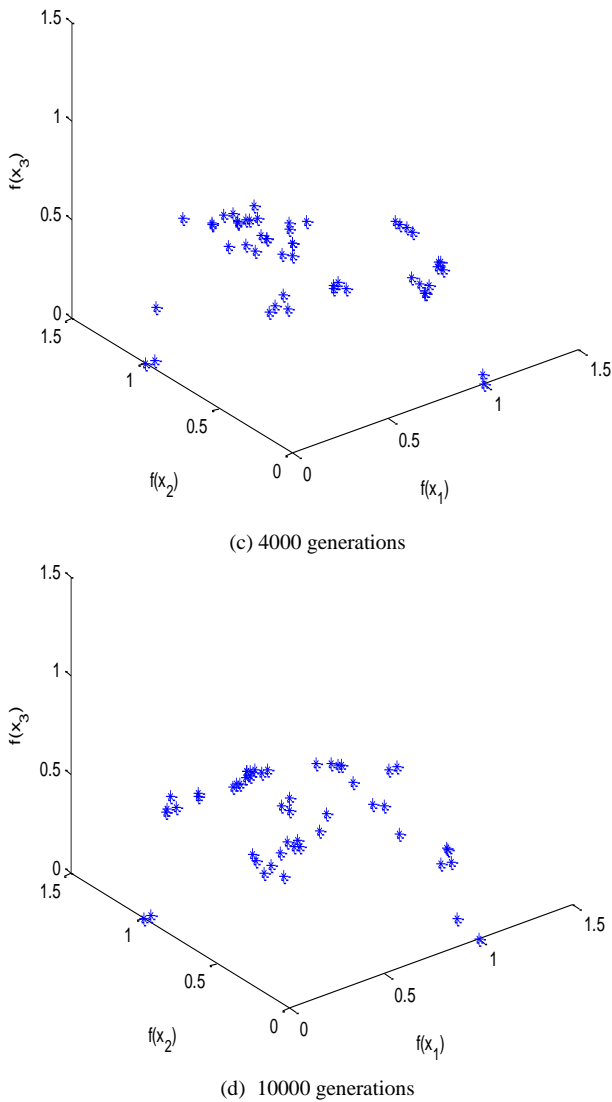

Fig. 11 Pareto solutions with different generations.

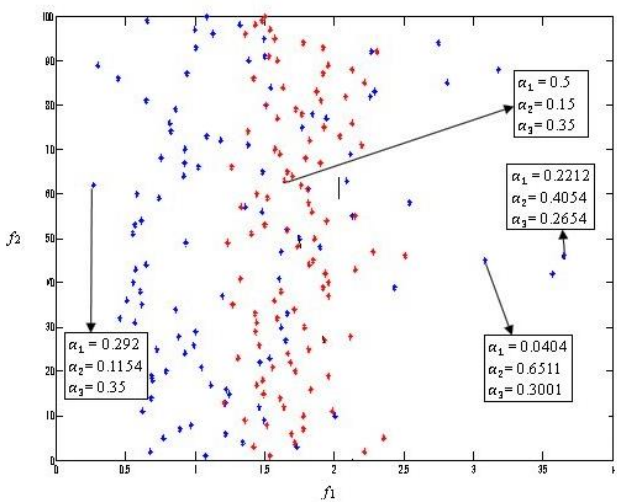

Fig.12 weighed-sum solution and Pareto solutions correlation. The points in red are the solution with $\alpha_1, \alpha_2, \alpha_3$ take values of 0.5, 0.15,0.35. while blue points are the solutions with random values of $\alpha_1, \alpha_2, \alpha_3$.

*D. Discussion*

There are fifty potential paths from the starting node (s) to the destination node (d) as an initial population. Six different selection methods are used in every generation in the suggested AGA. The extreme values of the fitness function are computed using these measures and are listed in Table 1. The best selection method is chosen for which the fitness function is the lowest because our work is of minimization type. The algorithm is ended when a maximum number of generations reached which is considered to be 100. From the above simulations, it can be seen that with the optimum path P4, the bandwidth is less than in the remaining paths, this is true because of the weighting factors that the WS approach adopts, where the weight $\alpha_2 = 0.15$ has been used for optimizing the bandwidth. On the other hand, the end-to-end delay and the minimum number of hops in the optimum path P4 were less than from the other paths, this is due to the high values of the weighting factors that have been used to optimize these objectives ($\alpha_1 = 0.5$ for the delay, $\alpha_3 = 0.35$ for the number of hops). To discuss how the WS approach can represent the Pareto Solutions, we proceed as follows, the general formula of the WS approach to MOGA is represented by (2). The values of $\alpha_1, \alpha_2, \alpha_3$ are chosen to increase the selection pressure on any of the two objective functions. In order to represent the Pareto technique, the weights $\alpha_1, \alpha_2, \alpha_3$ play a key role in this process. Which can take random values. So the WS approach can almost represents all solutions as shown in Figure 12. In this Figure the points in red color represent the solution produced by letting $\alpha_1, \alpha_2, \alpha_3$ take values of 0.5, 0.15,0.35 respectively, which are used in our design. The other points of blue color represent random values of $\alpha_1, \alpha_2, \alpha_3$. The total (red and blue) points represent the Pareto solutions. We conclude that the weighed-sum solution is a part of the whole possible Pareto solutions. Finally, When the results obtained from this work compared with another traditional technique, like dynamic programming techniques, we found that our proposed AGA performs better, where the total no. of hops and end-to-end delay obtained from our proposed AGA are 8 msec and 5 respectively as compared to 21 msec and 7 in [15–17,20].

## VII. CONCLUSIONS

Multi-objective *QoS* measures are achieved in this paper, these are: end-to-end delay, Bandwidth, and number of hops. Each one of them reflects the importance of a particular property in the wireless networks. To get better efficiency in the performance of WMN, we applied MOP to these objectives together. The suggested AGA is able of finding the solution for the multi-objectives wireless routing problems and gets the optimal solutions faster than the conventional procedures. The selection methods utilized in the suggested AGA of this work considerably had a direct effect on the behavior of the algorithm, it broadened the searching field and improved the procedure of the selection process. In the suggested algorithm, the proposed AGA adjusts quickly with changing networks environment, e.g., wireless networks with mobile nodes. On the other hand, concerning Pareto solutions, the diversity can be achieved as the number of generations increases, which makes the decision process easier at the higher level. Generally, they are not symmetric for higher dimensional spaces.

Acknowledgement

The Authors thanks Electrical engineering Department at the University of Baghdad for financial help and support.

REFERENCES

[1] H. Yetgin, K.T.K. Cheung, L. Hanzo, Multi-objective routing optimization using evolutionary algorithms, 2012 IEEE Wirel. Commun. Netw. Conf. (2012) 1–6. doi:10.1109/WCNC.2012.6214324.

[2] and P.K.G.T. Subarno Banerjee, Rajarshi Poddar, A Real Time Framework of Multiobjective Genetic Algorithm for Routing in Mobile Networks Subarno, ACEEE Int. J. Netw. Secur. 4 (2013) 11–15.

[3] Y. Donoso, R. Fabregat, J.L. Marzo, A multi-objective optimization scheme for multicast routing: A multitree approach, Telecommun. Syst. 27 (2004) 229–251. doi:10.123/B:TELS.0000041010.28247.5.

[4] E.G. N. Chase, M. Rademacher, A Benchmark Study of Multi-Objective Optimization Methods, 2009.

[5] J.H. Ryu, S. Kim, H. Wan, Pareto front approximation with adaptive weighted sum method in multiobjective simulation optimization, in: Proc. - Winter Simul. Conf., 2009: pp. 623–633. doi:10.1109/WSC.2009.5429562.

[6] A. Barolli, E. Spaho, L. Barolli, F. Xhafa, M. Takizawa, QoS routing in ad-hoc networks using GA and multi-objective optimization, in: Mob. Inf. Syst., 2011: pp. 169–188. doi:10.3233/MIS-2011-0116.

[7] M. Camelo, C. Omaña, H. Castro, QoS routing algorithms based on multi-objective optimization for mesh networks, IEEE Lat. Am. Trans. 9 (2011) 875–881.

[8] S. Rezaei, A.M.A. Hemmatyar, General study of jitter mechanisms for metric-based wireless routing protocols, AEU - Int. J. Electron. Commun. 79 (2017) 132–140. doi:10.1016/j.aeue.2017.05.044.

[9] M. Sepulcre, J. Gozalvez, B. Coll-Perales, Multipath QoS-driven routing protocol for industrial wireless networks, J. Netw. Comput. Appl. 74 (2016) 121–132. doi:10.1016/j.jnca.2016.08.008.

[10] M. Boushaba, A. Hafid, M. Gendreau, Node stability-based routing in Wireless Mesh Networks, J. Netw. Comput. Appl. 93 (2017) 1–12. doi:10.1016/j.jnca.2017.02.010.

[11] J. Wang, Y. Miao, P. Zhou, M.S. Hossain, S.M.M. Rahman, A software defined network routing in wireless multihop network, J. Netw. Comput. Appl. 85 (2017) 76–83. doi:10.1016/j.jnca.2016.12.007.

[12] T.S. Yuan Chai, Wenxiao Shi, Load-aware cooperative hybrid routing protocol in hybrid wireless mesh networks.pdf, (2017).

[13] Y. Rao, R. Wang, Performance of QoS routing using genetic algorithm for Polar-orbit LEO satellite networks, AEU - Int. J. Electron. Commun. 65 (2011) 530–538. doi:10.1016/j.aeue.2010.08.008.

[14] S.M.S. Alwan Nuha A S, Ibraheem K. Ibraheem, Fast Computation of the Shortest Path Problem through Simultaneous Forward and Backward Systolic Dynamic Programming, Int. J. Comput. Appl. 54 (2012) 21–25.

[15] Shukr Sabreen M., N.A.S. Alwan, I.K. Ibraheem, The Multi-Constrained Dynamic Programming Problem in View of Routing Strategies in Wireless Mesh Networks, Int. J. Inf. Commun. Technol. Res. 2 (2012) 471–476.

[16] I.K.I. Shukr, Sabreen Mahmood, Nuha Abdul Sahib Alwan, A Comparative Study of Single-Constraint Routing in Wireless Mesh Networks Using Different Dynamic Programming Algorithms Sabreen, J. Eng. 20 (2014) 49–60.

[17] A.A.A. Ibraheem Kasim Ibraheem, Application of an Evolutionary Optimization Technique to Routing in Mobile Wireless Networks Application of an Evolutionary Optimization Technique to Routing in Mobile Wireless Networks, Int. J. Comput. Appl. 99 (2016) 24–31. doi:10.5120/17385-7922.

[18] Z. Michalewicz, Genetic Algorithms + Data Structures = Evolution Programs, 1996.

[19] M. Mitchell, An introduction to genetic algorithms, Comput. Math. with Appl. 32 (1996) 133. doi:10.1016/S0898-1221(96)90227-8.

[20] I.K. Ibraheem, A.A.A. Ibraheem Kasim Ibraheem, Design of a Double-objective QoS Routing in Dynamic Wireless Networks using Evolutionary Adaptive Genetic Algorithm, Int. J. Adv. Res. Comput. Commun. Eng. 4 (2015). doi:10.17148/IJARCCE.2015.4935.

[21] T. Lu, J. Zhu, Genetic algorithm for energy-efficient QoS multicast routing, IEEE Commun. Lett. 17 (2013) 31–34. doi:10.1109/LCOMM.2012.112012.121467.

[22] C.W. Ahn, R.S. Ramakrishna, A genetic algorithm for shortest path routing problem and the sizing of populations, IEEE Trans. Evol. Comput. 6 (2002) 566–579. doi:10.1109/TEVC.2002.804323.

[23] A. Konak, D.W. Coit, A.E. Smith, Multi-objective optimization using genetic algorithms: A tutorial, Reliab. Eng. Syst. Saf. 91 (2006) 992–1007. doi:10.1016/j.ress.2005.11.018.

[24] J.D. Schaffer, Multiple objective optimization with vector evaluated genetic algorithms, 1st Int. Conf. Genet. Algorithms. (1985) 93–100. http://dl.acm.org/citation.cfm?id=657079.

[25] D.E. Goldberg, Genetic Algorithms in Search, Optimization, and Machine Learning, 1989. doi:10.1007/s10589-009-9261-6.

[26] M.G. PaulRani.A, PG, Multiobjective *QoS* Optimazation Based On Multiple Workflow

Scheduling In Cloud Environment, Int. J. Innov. Res. Comput. Commun. Eng. Vol. (n.d.).
[27] N. Srinivas, K. Deb, Muiltiobjective Optimization Using Nondominated Sorting in Genetic Algorithms, Evol. Comput. 2 (1994) 221–248. doi:10.1162/evco.1994.2.3.221.
[28] E.K. Burke, K. Graham, Search methodologies: Introductory tutorials in optimization and decision support techniques, second edition, 2014. doi:10.1007/978-1-4614-6940-7.

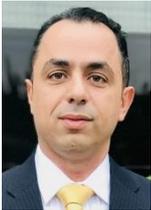

**Ibraheem K. Ibraheem** was born on August 23, 1976 in Baghdad, Iraq. He received the B.S. degree in electrical engineering from the university of Baghdad, Baghdad, Iraq in 1998 and M.Sc. & Ph.D degrees in Computer & Control Engineering from the same university and department, in 2001 and 2007, respectively. His research interests include Optimization in wireless networks, evolutionary and dynamic Optimizations, power control, Robotics, signal processing, nonlinear control and intelligent control applications.

**Alyaa A. Alhussainy** was born in Baghdad, Iraq, in 1979. She received the B.S. degree in hardware & software engineering from the university of Al-Mustasiriya, Baghdad, Iraq in 2001 and M.Sc. degree in Computer & Control Engineering from department of electrical engineering at Baghdad university in 2014. Her fields of interest includes wireless routing, evolutionary optimization, and networking protocols.